\pdfoutput=1

\documentclass[a4paper,10pt]{article}

\usepackage[utf8]{inputenc}
\usepackage{amssymb}
\usepackage{amsmath}
\usepackage{graphicx}
\usepackage{tabularx}
\usepackage{tabulary}
\usepackage{setspace}
\usepackage{authblk}
\usepackage[T1]{fontenc}
\usepackage[normalem]{ulem}
\usepackage{rotating}
\usepackage{color}
\usepackage[justification=centering,singlelinecheck=false,labelfont=bf,font=it]{caption}
\usepackage{natbib}

\title{\textbf{Linking 1D Evolutionary to 3D Hydrodynamical Simulations of Massive Stars}}
\author{\textbf{\uline{A.Cristini}$^a$, C.Meakin$^{b,c,d}$, R.Hirschi$^{a,e}$, D.Arnett$^b$, C.Georgy$^{a,f}$, M.Viallet$^g$}}
\affil{$^a$\textit{Astrophysics group, Keele University, Lennard-Jones Labs, Keele, ST5 5BG, UK}\\
$^b$\textit{Department of Astronomy, University of Arizona, Tucson, AZ 85721, USA}\\
$^c$\textit{New Mexico Consortium, Los Alamos, NM 87544, USA}\\
$^d$\textit{Theoretical Division, Los Alamos National Laboratory, Los Alamos, NM 87545, USA}\\
$^e$\textit{Kavli IPMU (WPI), The University of Tokyo, Kashiwa, Chiba 277-8583, Japan}\\
$^f$\textit{Geneva Observatory, University of Geneva, 1290 Versoix, Switzerland}\\
$^g$\textit{Max-Planck-Institut für Astrophysik, Karl Schwarzschild Strasse 1, Garching, D-85741, Germany}\\
\textit{Email: a.j.cristini@keele.ac.uk}}
\date{}

\bibpunct{(}{)}{;}{a}{,}{,}
\begin{document}

\maketitle

\abstract{\noindent Stellar evolution models of massive stars are important for many areas of astrophysics, for example nucleosynthesis yields, supernova progenitor models and understanding physics under extreme conditions. Turbulence occurs in stars primarily due to nuclear burning at different mass coordinates within the star. The understanding and correct treatment of turbulence and turbulent mixing at convective boundaries in stellar models has been studied for decades but still lacks a definitive solution. This paper presents initial results of a study on convective boundary mixing (CBM) in massive stars. The ‘stiffness’ of a convective boundary can be quantified using the bulk Richardson number ($\textrm{Ri}_B$), the ratio of the potential energy for restoration of the boundary to the kinetic energy of turbulent eddies. A ‘stiff’ boundary ($\textrm{Ri}_B \sim 10^4$) will suppress CBM, whereas in the opposite case a ‘soft’ boundary ($\textrm{Ri}_B \sim 10$) will be more susceptible to CBM. One of the key results obtained so far is that lower convective boundaries (closer to the centre) of nuclear burning shells are ‘stiffer’ than the corresponding upper boundaries, implying limited CBM at lower shell boundaries. This is in agreement with 3D hydrodynamic simulations carried out by \citet{2007ApJ...667..448M}. This result also has implications for new CBM prescriptions in massive stars as well as for nuclear burning flame front propagation in Super-Asymptotic Giant Branch (S-AGB) stars and also the onset of novae.}

\clearpage

\noindent \textit{This work was presented at the ``Turbulent Mixing and Beyond'' 2014 workshop as a contributed talk. It tackles one of the target problems of the workshop: the evolution of fluid boundaries and convection in fluids. Also, this work answers questions within one of the themes of the workshop: determining momentum transfer in buoyancy driven convective flows. Both of these phenomena are explored within an astrophysical setting.}\\ 

\section{An Overview of Massive Stars}

\noindent Massive stars play a key role in the universe through the light they shine, the elements they produce and the explosion that marks their death. Consequently, their evolution and structure are important for many other astrophysical phenomena: nucleosynthesis, pre-supernova progenitors, stellar populations and compact objects are all dependant.\\

\noindent Massive stars are those with an initial mass large enough to allow them to initiate core-collapse at the end of their lives. According to \citet{2003fthp.conf....3H,2002RvMP...74.1015W} and works since this workshop by \citet{2015MNRAS.447.3115J} the mass limit is around 8$M_\odot$.
 For a star without rotation, magnetic fields or close companions the only forces are due to gravity and an internal pressure gradient. For most of the star's life these two forces compensate each other and the star is said to be in hydrostatic equilibrium. The star's energy source comes from the nuclear fusion of lighter nuclei into heavier ones. This process begins in the core with fusion of hydrogen nuclei to produce helium. As temperatures and densities rise eventually helium nuclei fuse together to create carbon. This process repeats, burning carbon, neon\footnote{Oxygen cannot be formed from fusion of neon nuclei, as neon is heavier. Instead neon is broken up into an oxygen and helium 
nucleus by an energetic photon, this process is known as photodisintegration.}, oxygen and silicon, consecutively in the core. In the final burning stage iron is produced from a quasi-equilibrium of alpha-captures ($^A\textrm{X}+\;\alpha\rightarrow\;^{A+4}\textrm{Y}$) and photodisintegration, starting with the seed nuclei $\,^{28}\textrm{Si}$.\\

\noindent Once iron is produced in the core fusion cannot proceed as the peak in binding energy per nucleon is reached. With no energy generation from nuclear burning in the core, eventually the pressure gradient is not large enough to keep the star in hydrostatic equilibrium. Temperatures and densities rise as the core collapses eventually leading to a bounce and then a supernova explosion. Almost all of the stellar material is ejected, with either a neutron star or black hole remnant remaining (dependent upon the initial mass of the star).\\

\noindent In massive stars, in addition to energy sources from nuclear burning and gravitational contraction, there can be energy losses due to radiation from the surface (luminosity) and neutrinos produced in the stellar plasma. These neutrinos dominate energy losses over radiation in between helium and carbon burning \citep{1996snih.book.....A}. Since neutrinos emitted in the core during the advanced phases can freely escape from the star (their mean-free path being larger than the star until much higher densities are reached), the evolution of massive stars accelerates and photon energy losses play a negligible role in the inner regions for the rest of the evolution.\\

\noindent Such high temperatures in stellar interiors leads to a highly ionised medium, the state of matter in the star is therefore a plasma. The plasma has a very small Knudsen number\footnote{The Knudsen number is the ratio of the mean free path to a representative length scale.} ($\sim10^{-15}$) implying that there are many collisions between particles; and can therefore be modelled as a continuous fluid \citep{1960trht.book..220P}. The Reynolds number\footnote{The Reynolds number is the ratio of inertial forces to viscous forces.} ($\sim10^{12}$) indicates that this fluid is highly turbulent in convective zones. Since nuclear reactions have a steep temperature dependence, energy production peaks are narrow in radius and such a high flux in energy leads to convective instability.\\

\noindent 1D stellar models can be used to model a star from birth to death. Due to the complexity of stars simplifications must be made before their structure can be calculated. Stars are assumed to be spherically symmetric, collapsing the stellar evolution equations (see Eqs. \ref{stellar} - \ref{stellar1}) into 1D, allowing them to be solved along the radius, although it is more common to solve them within a Lagrangian framework. Stellar models can be calculated for any mass and metallicity (the percentage of elements heavier than He), with the inclusion of more complex physics such as rotation, magnetic fields and binarity. These models are valuable in many other areas of astrophysics such as the study of galactic chemical evolution, galactic dynamical evolution (for which turbulence plays a major role in) and the properties of planet hosting stars.\\

\noindent A major uncertainty in stellar evolution models is the treatment of chemical mixing in convectively unstable regions and their adjacent boundary regions. Stellar models have evolved to use parameterisations for chemical mixing at the boundaries \citep{1991A&A...252..179Z,1996A&A...313..497F}, these are based on results from 3D hydrodynamical simulations. Simulations of the oxygen burning shell in a 23 M$_{\odot}$ star \citep{2007ApJ...667..448M} show that material is mixed into the convective region through turbulent entrainment on a timescale less than the crossing time of convective eddies. Furthermore, their rate of entrainment is in agreement with geophysical and atmospheric science literature \citep[e.g.][]{2004JAtS...61..281F}, and follows a scaling law that is inversely proportional to the boundary `stiffness'.\\ 

\noindent In this paper we present a parameter study of various convectively unstable regions that develop during a star's evolution modelled using a 1D stellar evolution code. This study will inform us on which stages of stellar evolution are best suited to compressive reactive-hydrodynamic simulations within the Implicit Large Eddy Simulation (ILES; \citealt{2006JTurb...7N..15M}) paradigm. We must be able to simulate multiple convective cycles to allow for a statistically valid study, for this we need a reasonable amount of computing time. 

\section{1D Models of Massive Stars}

\subsection{Method and Numerical Tool}

\noindent 1D modelling of massive stars provides an insight into the evolution of stars from the main sequence to core collapse. As given on Pg. 89 of \citet{2013sse..book.....K} the full set of stellar evolution equations, in Lagrangian form are:

\begin{align}\label{stellar}
\frac{\partial r}{\partial m}&=\frac{1}{4\pi r^2\rho};\\
\frac{\partial P}{\partial m}&=-\frac{Gm}{4\pi r^4};\\
\frac{\partial L}{\partial m}&=\epsilon_n - \epsilon_{\nu}-c_P\frac{\partial T}{\partial t}+\frac{\delta}{\rho}\frac{\partial P}{\partial t};\\
\frac{\partial T}{\partial m}&=-\frac{GmT}{4\pi r^4P}\triangledown;\\
\frac{\partial X_i}{\partial t}&=\frac{m_i}{\rho}\left(\sum_jr_{ji}-\sum_kr_{ik}\right),\:\:\:i=1,...,I;
\label{stellar1}
\end{align}

\noindent where $\epsilon_{n,\nu}$ is the energy generation per unit mass per second due to nuclear reactions and neutrinos, respectively; $\triangledown=\left(\frac{\partial\,\mathrm{ln}\,T}{\partial\,\mathrm{ln}\,P}\right)$; $\delta=-\left(\frac{\partial\,\mathrm{ln}\,\rho}{\partial\,\mathrm{ln}\,T}\right)$; $X_i$ is the mass fraction of nuclear species $i$ with mass $m_i$; $r_{ab}$ is the reaction rate for nuclear species $a\rightarrow b$; the remaining variables have their usual meaning.\\

\noindent The above stellar evolution equations can be implicitly solved using a finite differencing scheme. Variables $r,P,L,T \;\mathrm{and}\; X_i$ can be obtained given closure from: the equation of state (EOS), $P(\rho,T,X_i)$; specific heat capacity at constant pressure, $c_P$; opacity, $\kappa$; thermodynamic exponent $\delta$; reaction rate $r_{jk}$; and energy generation rates $\epsilon_{n,\nu}$. In radiative regions the temperature gradient, $\triangledown$, can be replaced by the radiative temperature gradient $\triangledown_{rad}=\frac{3}{16\pi acG}\frac{mT^4L}{\kappa P}$. For convective regions $\triangledown$ can be calculated using a theory of convection, e.g. mixing length theory, or assumed to be adiabatic, $\triangledown_{ad}=\frac{P\delta}{T\rho\; c_P}$.\\

\noindent The Geneva Stellar Evolution Code (GENEC, \citealt{2008Ap&SS.316...43E}), which we have used to calculate the models in this study, can model stars of various initial masses (sub-solar to very massive) at different metallicities. Massive stars can be modelled from the zero age main sequence to the end of silicon burning.\\

\noindent Multi-dimensional processes such as: rotation; magnetic fields; mass loss due to stellar winds; advection (the bulk motion of matter); and additional chemical mixing are parameterised into 1D. The abundances of 23 isotopes\footnote{The complete list of isotopes followed in GENEC are: $^1$H, $^3$He, $^4$He, $^{12}$C, $^{13}$C, $^{14}$N, $^{15}$N, $^{16}$O, $^{17}$O, $^{18}$O, $^{20}$Ne, $^{22}$Ne, $^{24}$Mg, $^{25}$Mg, $^{26}$Mg, $^{28}$Si, $^{32}$S, $^{36}$Ar, $^{40}$C, $^{44}$Ti, $^{48}$Cr, $^{52}$Fe, $^{56}$Ni} in the range $^1$H to $^{56}$Ni are calculated during the evolution. The many instabilities associated with rotation are also included (see \citealt{2004A&A...425..649H, 2010A&A...522A..39F} for more details). Differential rotation in radiative regions produces anisotropic turbulence. Turbulence is strong along isobars, which leads to constant angular velocity on isobaric surfaces \citep{2009pfer.book.....M}. This means that in stellar models rotation can be treated as shellular, where the angular velocity is constant along isobars \citep{1992A&A...265..115Z}.\\

\noindent The EOS used in GENEC considers an ensemble of a perfect gas of ions, partial degeneracy of electrons and radiation from photons. For lower mass stars and in the envelopes of stars, partial ionization and coulomb effects must be considered, so the OPAL EOS \citep{1996ApJ...456..902R} and the Mihalas-Hummer-Dappen EOS (\citealt{1988ApJ...331..794H, 1988ApJ...332..261D}) are used in these situations. Magnetic fields are generated by turbulent motion and rotation, they are modelled in GENEC through the Taylor-Spruit dynamo \citep{2002A&A...381..923S}. The following are also included in the code: mass loss; advection; penetrative overshooting; rotationally induced chemical mixing; and angular momentum transfer (see \citealt{2010A&A...522A..39F} and references therein for more details).

\subsection{Treatment of Convection}

\noindent Convection, in the simplest sense, can be described by considering a local bubble within a gas. The bubble is acted on by an upward buoyant force and a downward force due to its weight. If these forces are imbalanced in either direction the bubble will be displaced. In its new position the forces on the bubble can now have a restoring effect, for example, if the bubble is displaced upwards but is now heavier than the surrounding gas it will travel back towards its original position. While in the opposite case after being displaced the forces on the bubble could be equally or more unbalanced, for example, if the bubble is displaced upwards and is lighter than the surrounding fluid it will continue to rise, this is the dynamically unstable case of convection.\\

\noindent Two criteria exist to predict locally if a medium will be convective, they are distinguished by their consideration of the composition and whether it varies throughout the medium. For a mixture of nuclei of type $i$ a mean molecular weight can be calculated by:\\

\begin{equation}\label{mu}
\mu=\left(\sum_i\frac{X_i(1+Z_i)}{\mu_i}\right)^{-1}=\frac{\rho}{m_u}\left(\sum_in_i(1+Z_i)\right)^{-1},
\end{equation}

\noindent for molecular weight, $\mu_i$ and charge number, $Z_i$. A region is locally and dynamically unstable due to convection if:\\

\begin{equation}\label{ledoux}
 \triangledown_{rad}>\triangledown_{ad}+\frac{\phi}{\delta}\triangledown_\mu,
\end{equation}

\noindent where $\triangledown_{\mu}=\frac{\partial\textrm{ln}\mu}{\partial\textrm{ln}P}$ and $\phi=\left(\frac{\partial\,\mathrm{ln}\,\rho}{\partial\,\mathrm{ln}\,\mu}\right)_{P,T}$.
If the chemical composition effects are ignored then Eq. \ref{ledoux} simplifies to the Schwarzchild criterion for convective instability:\\

\begin{equation}\label{schw}
 \triangledown_{rad}>\triangledown_{ad}.
\end{equation}

\noindent In the case of a region being unstable according to the Schwarzchild criterion but stabilised by a molecular weight gradient (i.e. stable according to the Ledoux criterion), it can be described as semi-convective. If within a semi-convective region the mixing is assumed to be strong enough, then the stabilising molecular weight gradient will be washed out and the region can be considered fully convective \citep{2004PhDT........RH}. In this case the simpler Schwarzchild criterion can be used to determine the local extent of convection. For most stellar evolution codes the Schwarzchild criterion is used. The inclusion of the Ledoux criterion requires a theoretical prescription for semi-convection.\\

\noindent The most commonly used prescription for convection in stellar models is the mixing length theory (MLT). 
MLT, developed by \citet{1958ZA.....46..108B} can be used to calculate the temperature gradient and velocity within a convective region. The range of sizes, velocities, densities and temperatures of convective elements are all averaged out using a single length scale, representative of the average of these physical quantities. Within a convective region, bubbles will rise vertically over an average distance known as the mixing length, $\ell=\alpha H_P$, where $H_P$ is the pressure scale height defined as,\\

\begin{equation}\label{hp}
 H_P = -\frac{dr}{d\,\mathrm{ln}P}\;,
\end{equation}

\noindent the free parameter $\alpha$ must be constrained by observations, in order for the theory to be able to produce similar convective velocities to stellar interiors. Beyond the mixing length bubbles are assumed to thermalise with the surrounding fluid and dissipate their internal energy. All bubbles are assumed to have a radius of the mixing length. This assumption is not applicable to all convective regions, for example in the case of main sequence stars of mass $<10$ M$_\odot$ whose convective cores can be smaller than one pressure scale height \citep{1987A&A...188...49R}.
Bubbles are assumed to be in pressure equilibrium with their surroundings until they dissolve. Instead of a spectrum of temperature gradients an average is taken, above this the bubbles will rise and below they will sink. Radiative losses are accounted for so the bubbles do not quite rise adiabatically.\\

\noindent With MLT one can calculate the convective diffusion coefficient and therefore the strength of convection using the velocity and mixing length.
Other prescriptions for convection in stellar models exist such as turbulent convection models by \citet{1996ApJ...473..550C} and \citet{1991ApJ...370..295C}. These models have not yet been implemented into stellar evolution models, due to their complexity.\\

\noindent Convection is a multi-dimensional process and therefore must be prescribed or parameterised into one dimension, for use in stellar evolution models. For a convective shell sandwiched between two stable regions, convection is the main contributor to turbulent mixing. Convective velocities at the boundaries are non-zero giving rise to mixing or entrainment of material at the boundary. Buoyancy forces on a convective element heavier than its surroundings result in oscillations around an equilibrium position, these oscillations can be seen as internal gravity waves within the stable region \citep{LectureA}.\\

\noindent Turbulent entrainment and wave generation cannot be modelled in stellar evolution calculations, due to a simplified picture of convection given by MLT. Although MLT can predict the convective velocity it does not take into account the non-locality of the turbulent flow. A combination of local criterion for convection (Eqs. \ref{ledoux} and \ref{schw}) and `overshooting' prescriptions can provide an estimate for additional mixing. Two commonly adopted prescriptions for the strength of mixing beyond locally defined boundaries are penetrative \citep{1991A&A...252..179Z} and diffusive \citep{1996A&A...313..497F} overshooting, these methods account for convective eddies with non-zero velocities beyond the boundary.\\\\

\begin{figure}[h!]
\centering
\includegraphics[width=0.85\textwidth]{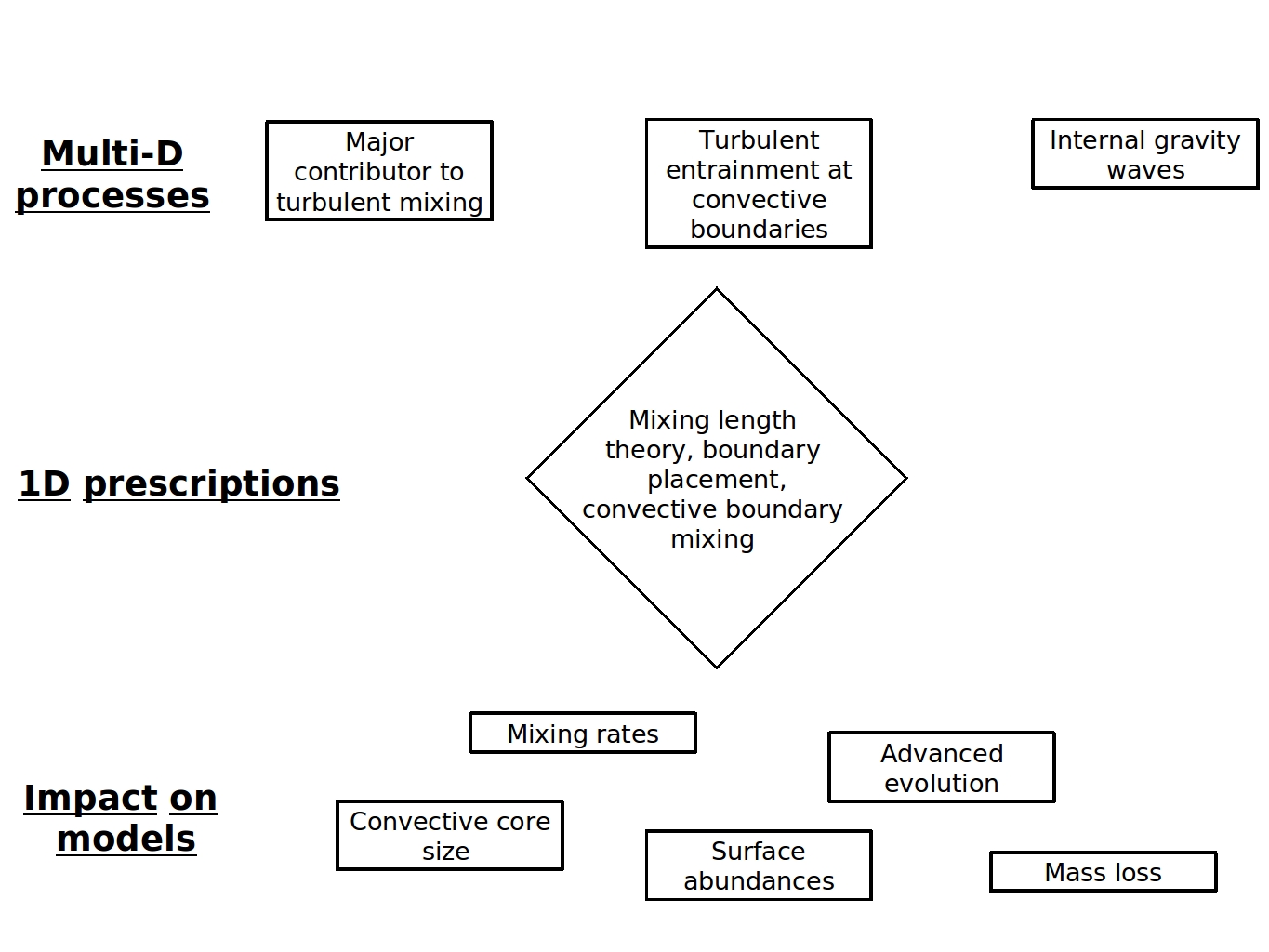}
\caption{Schematic of 3D processes, 1D modelling prescriptions and the consequential impacts on stellar models.}
\label{presc}
\end{figure}

\noindent Fig. \ref{presc} illustrates how stellar evolution calculations can be sensitive to these prescriptions for convection, boundary placement and convective boundary mixing. Despite most stellar evolution codes using MLT, the plethora of treatments and prescriptions for convection result in large differences between similar models using different codes \citep{2013A&A...560A..16M}. \citet{2014MNRAS.439L...6G} also show that the use of either the Ledoux (Eq. \ref{ledoux}) or Schwarzschild (Eq. \ref{schw}) criteria can lead to large differences in model evolution. There is room to better constrain these 1D prescriptions to improve stellar models; one solution is to use 3D simulations.

\section{3D Models of Massive Stars}
\subsection{Method and numerical tool}
\noindent Currently, the modelling of turbulence using 1D stellar evolution codes is inadequate. In addition to this, these models cannot simultaneously, self-consistently model energy transport and mass entrainment. Hence, there is a clear need for multi-D simulations of stellar convection to either, 1) provide new self-consistent prescriptions for convection in stellar models, or 2) constrain free parameters used in current prescriptions for convection in stellar models. Multi-D simulations allow one to study from first principles the process of convective boundary mixing, consequently much shorter time scales are used than in 1D stellar evolution modelling. This requires shorter time steps, greatly increasing the cost of such simulations. 3D hydrodynamical simulations are therefore limited to sub-regions of stars and only for a small fraction of their lifetime, as simulations of full stars are too computationally expensive. Studying fluid dynamics using 3D hydrodynamics can improve the treatment and prescriptions of turbulence in 1D stellar models, and therefore achieve more accurate evolutionary paths for stars. In turn, improved 1D stellar evolution models can aid in improving their application in other areas of astrophysics, such as nucleosynthesis yields, pre-supernova progenitor models and stellar populations.\\

\noindent 3D reactive hydrodynamic simulations can be used to model turbulence and turbulent mixing in stellar interiors. In our approach we solve the Euler equations for a compressible, inviscid flow including composition change due to nuclear reactions:

\begin{align}
\frac{\partial\rho}{\partial t} + \boldsymbol{\nabla}\cdot(\rho \mathbf{u}) &= 0\\
\rho\frac{\partial \mathbf{u}}{\partial t} + \rho\mathbf{u}\cdot\boldsymbol{\nabla}\mathbf{u} &= -\boldsymbol{\nabla}p +\rho\mathbf{g}\\
\rho\frac{\partial E}{\partial t} + \rho\mathbf{u}\cdot\boldsymbol{\nabla}E + \boldsymbol{\nabla}\cdot(p\mathbf{u})&=\rho\mathbf{u}\cdot\mathbf{g}+\rho(\epsilon_{n}+\epsilon_\nu)\\
\rho\frac{\partial X_i}{\partial t} + \rho\mathbf{u}\cdot\boldsymbol{\nabla}X_i &= R_i
\end{align}

\noindent where $E$ is the total specific energy and $R_i$ is the rate of change of composition due to nuclear burning.\\ 

\noindent Our numerical calculations are performed using the hydrodynamics code PROMPI, within the ILES paradigm. ILESs focus on resolving the largest eddies, and avoid the use of a sub-grid scale model for the dissipation of kinetic energy below the grid scale. Instead these simulations use the truncation errors due to the discretisation of the problem to act as a physically motivated sub-grid scale model \citep{2014PhFl...26j6101H}. ILESs rely on this assumption to hold for very large Reynolds numbers. Scales at which dissipation occurs can be modelled through direct numerical simulations (DNS), but such simulations cannot scale up to stellar lengthscales within the current era of computational power. Further work since this workshop by \citet{2015ApJ...809...30A} discusses in detail these two types of simulations.\\ 

\noindent PROMPI is a MPI parallelised version of a finite-volume, Eulerian, piecewise parabolic method (PPM) implementation of \citet{1984JCoPh..54..174C} derived from the legacy astrophysics code PROMETHEUS \citep{1989nuas.conf..100F}. The base hydrodynamics solver is complemented by micro-physics to treat the equation of state and nuclear reactions, as well as self-gravity in the Cowling approximation \citep[pg. 86 of][]{2000itss.book.....P} appropriate for deep interiors.

\subsection{Insight from 3D simulations}

\noindent Our approach to the multi-D modelling of massive stellar interiors follows the work of \citet{2007ApJ...667..448M, 2009ApJ...690.1715A}; and \citet{2013ApJ...769....1V}. A non-exhaustive list of other hydrodynamical simulations for lower mass stars includes \citet{2014ApJ...792L...3H,2012rgps.book...87M,2011ApJ...742..121S,2010ASPC..429..167V}; and \citet{2009A&A...501..659M}. \\

\noindent \citet{2007ApJ...667..448M} used the multi-dimensional hydrodynamics solver PROMPI to calculate 3D simulations of hydrogen core burning and oxygen shell burning of a 23M$_\odot$ star.
They concluded that the MLT prescription for convection is incorrect, due to its 
assumption that the net up/down-flowing kinetic energy is zero.
They also found that convective boundary mixing is more complicated and varied than prescriptions currently used in the literature,  
for example, the overshooting phenomenon is more akin to an elastic reaction of the boundary due to fluid motions. In fact, it was shown that mixing across a boundary actually occurs in a boundary region, where convective eddies are decelerated.\\ 

\noindent Within the fluid dynamics and atmospheric science literature mixing is seen to occur through turbulent entrainment. Entrainment is the transport of fluid across an interface between two bodies of fluid by a shear induced turbulent flux \citep{1973Sci...180.1356T}. Entrainment of material at the boundary is inversely proportional to the buoyancy jump \citep{2007ApJ...667..448M}, which is in agreement with studies on the planetary boundary layer (e.g. \citealt{1998JAtS...55.3042S}). For example, the amount of mixing and the timescales at which they occur may therefore be quite different from penetrative overshooting.\\

\begin{figure}[h!]
\centering
\includegraphics[width=0.8\textwidth]{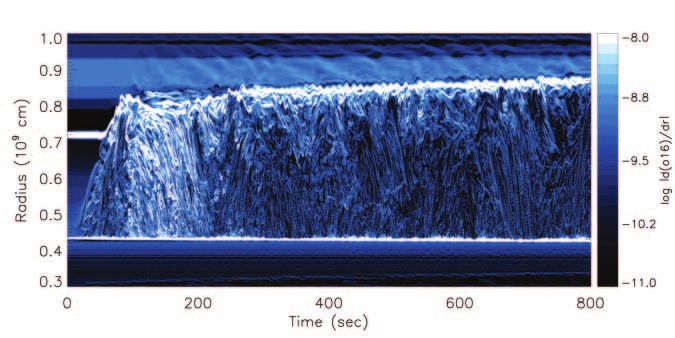}
\includegraphics[width=0.8\textwidth]{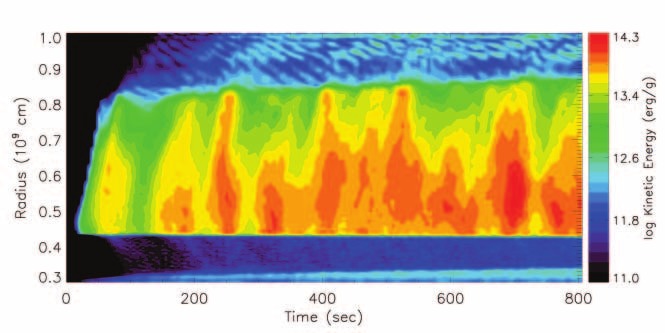}
\caption{From \protect\citet{2007ApJ...667..448M}. 3D hydrodynamical simulation of oxygen shell burning in a 23 M$_\odot$ star with $\sim$60\% of fuel burnt. Top: Colours show the oxygen abundance gradient. Bottom: Colours show the specific turbulent kinetic energy of the fluid. Convective boundary mixing clearly occurs around the upper convective boundary, indicated in the top panel by the thick white line.}
\label{3Dcbm2}
\end{figure}

\noindent The top panel of Fig. \ref{3Dcbm2} shows initially a static lower and upper convective-radiative boundary. The upper boundary at a radius of $0.7\times10^9$\,cm is described by the Ledoux criterion (Eq. \ref{ledoux}). Convection is initiated through a random low amplitude perturbation in temperature and density ($\sim 0.1\%$). Turbulent motions are observed as the region becomes convectively unstable between $0.43\times10^9$\,cm and $0.7\times10^9$\,cm. Surrounding this zone on either side are convectively stable radiative zones. Penetration of the upper convective boundary develops over time in a cumulative manner and into a considerable fraction of the shell radius. Fuel is entrained into the burning region, this changes the energy generation and structure of the shell. \citet{2007ApJ...667..448M} use the bulk Richardson number (see Eq. \ref{ribeq}) as a diagnostic tool for the amount of mixing beyond the boundary which was originally adopted in the atmospheric science field.\\

\begin{figure}[h!]
\centering
\includegraphics[width=0.85\textwidth]{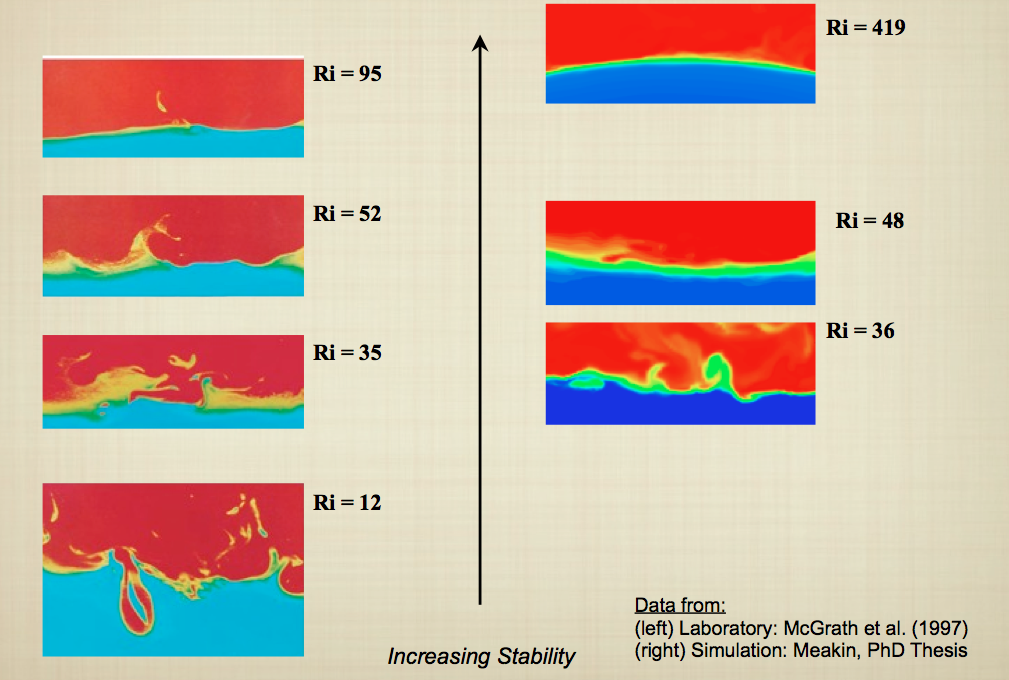}
\caption{From \protect\citet{2006PhDT........CM}. Left: water tank experiments at different values of bulk Richardson number. Right: spherical hydrodynamic simulations for different values of bulk Richardson number.}
\label{rib}
\end{figure}

\noindent Examples of compositional mixing across convective boundaries are demonstrated in single-sided stirring water tank experiments by \citet{1997JFM...347..235M} and hydrodynamical simulations in a spherical geometry \citep{2006PhDT........CM}, shown in Fig. \ref{rib}.
For increasing values of the bulk Richardson number (Ri$_B$, Eq. \ref{ribeq}), different mixing processes occur. For low values ($\textrm{Ri}_B \lesssim 15$) mixing occurs through the rebounding and impingement of plumes at the interface. Intermediate values ($15 < \textrm{Ri}_B \lesssim 35$) shear mixing at the interface occurs due to plumes with high horizontal velocities. Finally at higher values (Ri$_B > 35$) the presence of interfacial waves and their breaking events are the dominant form of mixing.\\

\noindent The physical regime for stellar plasmas is an extreme one, as shown in Table \ref{tab1}. The typical Reynolds number\footnote{Using data from a stellar model of the carbon shell burning phase of a 15M$_\odot$ star.} is $10^{12}$, while the largest values obtained in the laboratory are $\sim10^7$ \citep{2002MeScT..13.1608M}. Turbulence typically develops in laminar flows for Reynolds numbers greater than $10^3$, so one can see that stellar material is highly turbulent. To resolve all scales in a star from the dissipation scale to the advective scale of the largest eddies would require $\sim 10^{40}$ zones (see works since this workshop by \citealt{2015ApJ...809...30A}). The latest simulations (\citealt{2014ApJ...792L...3H}) can reach $\sim 10^{10}$ zones, which is equivalent to a Reynolds number of $\sim 10^4$ \citep{1959flme.book.....L}.\\

\noindent Of course, convective boundary mixing and entrainment is not only important for stellar physics, another contributor to the workshop \citep[see][]{2014JPhCS.547a2042G} present results from DNS of a cloud - clear air interface  which develops into two turbulent regions separated by an interfacial region.\\

\noindent Turbulent entrainment of stellar interiors using 3D hydrodynamic simulations have only been conducted for a handful of stellar nuclear burning phases, and even less for massive stellar interiors. Hence, before new simulations are undertaken it is important to consider all of the important characteristics that describe the fluid and bulk properties. With this information it is possible to say: whether detailed `real-time' calculations with true luminosities can be done; whether any computational simplifications can be made (e.g. neglecting radiative diffusion due to efficient convection); and also how much of the evolution can be calculated given a set amount of computing time. To answer these questions we have conducted a parameter study on different phases of evolution of a 15 M$_\odot$ star.

\begin{table}
\captionsetup{justification=centering}
\hskip-9.0mm\begin{tabulary}{\textwidth}{c c c c c c c c c}
\hline \hline\\
\textbf{T} & $\boldsymbol{\rho}$ & \textbf{r$_{c}$} & \textbf{v} & $\boldsymbol{\nu}$ & $\boldsymbol{\kappa}$ & $\boldsymbol{\nu_T}$ & \textbf{g} & \textbf{D}\\\\
\hline\\
5$\times10^8$  &  \;\;2$\times10^4$  &  \;\;2$\times10^{9}$  &  \;\;2$\times10^5$  &  \;\;320\;  &  \;\;0.1\;\;  &  \;\;2$\times10^{6}$  &  \;\;7$\times10^7$  &  \;\;5$\times10^{13}$\\\\
\hline \hline\\
\textbf{Re} & \textbf{Pr} & \textbf{Pe} & \textbf{Ra} & \textbf{Ma} & \textbf{At} & \textbf{Sc} & \textbf{Kn} & \textbf{Ca}\\\\
\hline\\
\;\;1$\times10^{12}$  &  \;\;1$\times10^{-4}$  &  \;\;2$\times10^{8}$  &  \;\;1$\times10^{26}$  &  \;\;9$\times10^{-4}$ & \;\;5$\times10^{-3}$ & \;\;6$\times10^{-12}$ & \;\;1$\times10^{-15}$ & \;\;8$\times10^{-7}$\\\\
\hline\hline
\end{tabulary}
\caption{Structural and flow properties for the carbon burning shell of a 15M$_\odot$ stellar model.\\ 
Top row - Temperature (K), density (g\,cm$^{-3}$), convective shell radius (cm), turbulent velocity (cm\,s$^{-1}$), radiative viscosity (cm$^2$\,s$^{-1}$), opacity (cm$^2$\,g$^{-1}$), thermal diffusivity (cm$^2$\,s$^{-1}$), gravitational acceleration (cm\,s$^{-2}$) and convective diffusion coefficient (cm$^2$\,s$^{-1}$), respectively.\\
\textrm{Bottom row} - Reynolds, Prandtl, P\'eclet, Rayleigh, Mach, Atwood, Schmidt, Knudsen and Cauchy numbers, respectively.}
\label{tab1}
\end{table}

\section{Parameter Study of convective boundaries in massive stars}
\subsection{Theoretical framework}

\noindent In preparation for future multi-D simulations of convective boundaries in massive stars, a parameter study of various structural and flow properties of convective regions and the surrounding stable layers was conducted.\\ 

\noindent Various quantities were analysed over the convective and stable regions either side of the boundary: gravitational acceleration $g$; pressure scale height $H_P$; luminosity $L$; mean molecular weight $\mu$; convective velocity, $v_c$, which is estimated from the convective flux, $F_c$ and given by:\\

\begin{equation}
v_c=\left(\frac{F_c}{\rho}\right)^{\frac{1}{3}};
\end{equation}

\noindent the convective velocity predicted by MLT, given by:\\

\begin{equation}
v_{MLT}=\left(\frac{1}{4}\frac{L}{4\pi r^2}\frac{g \triangledown_{ad}\alpha H_P}{P}\right)^\frac{1}{3},
\end{equation}

\noindent with $\alpha=\left(\frac{\partial\, \mathrm{ln}\,\rho}{\partial\, \mathrm{ln}\, P}\right)_{\mu,T}$; the Mach number of the flow, given by:\\

\begin{equation}
\textrm{Ma}=\frac{v_c}{c_s}=\sqrt{\frac{\rho}{\gamma P}}\;v_c,
\end{equation}

\noindent where $\gamma$ is the ratio of specific heats; and the bulk Richardson number given as the ratio of `stabilisation potential' to turbulent kinetic energy and written as:\\

\begin{equation}
\textrm{Ri}_B=\frac{\Delta B\; \mathcal{L}}{v_c^2/2}.
\label{ribeq}
\end{equation}

\noindent The `stabilisation potential' ($\Delta B\; \mathcal{L}$) is akin to the work done by the boundary against convective motions. $\mathcal{L}$ is a length scale associated with the turbulent motions and $\Delta B$ is the buoyancy jump (see Eq. \ref{buoyj}) over some distance $\Delta r$ either side of the boundary position, $r_0$. For consistency we have kept the turbulent length scale $\mathcal{L}$ and the integration distance over the boundary $\Delta r$ of the same order, which is taken as 10$\%$ and 5$\%$ of the pressure scale height for convective cores and shells, respectively. The buoyancy jump is determined using the following formula:\\

\begin{equation}
\Delta B=\int\limits_{r_0-\Delta r}^{r_0+\Delta r}N^2dr,
\label{buoyj}
\end{equation}

\noindent where $\Delta B=\Delta B(N_T^2)+\Delta B(N_\mu^2)$, is the sum of its thermal and compositional components and $N$ is the Brunt-V\"{a}is\"{a}l\"{a} frequency.\footnote{The Brunt-V\"{a}is\"{a}l\"{a} frequency is given by $N=\sqrt{N_T^2+N_\mu^2}=\sqrt{\frac{g}{H_P}\left[\delta\left(\triangledown_{ad}-\triangledown\right)+\phi\triangledown_{\mu}\right]}$}\\

\noindent Note that $N^2$ is positive for stability and negative for instability due to convection, $N_T$ and $N_\mu$ represent the thermal and compositional components, respectively. $v_c^2/2$ is the specific turbulent kinetic energy within the convective region. 

\subsection{Results}

\begin{figure}[h!]
\centering
\includegraphics[width=0.85\textwidth]{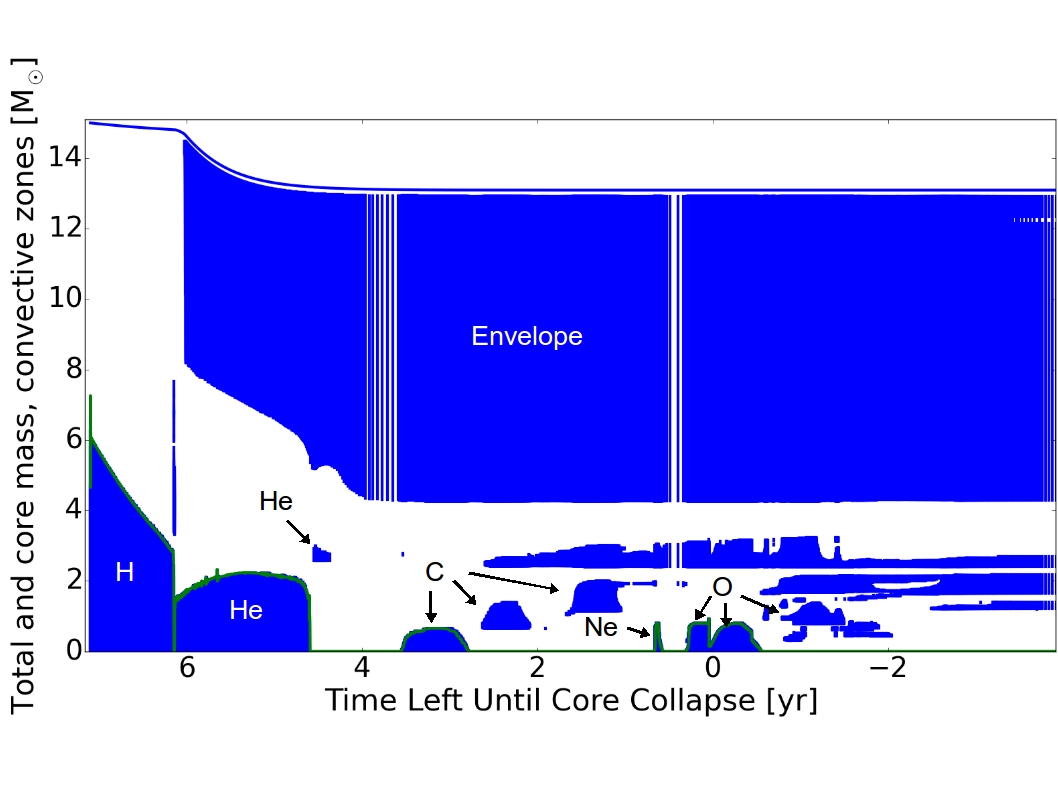}
\caption{Structure evolution diagram of a 15M$_\odot$ stellar model. Interior mass is plotted on the vertical axis and the log of the time left until collapse on the horizontal axis. Coloured areas represent convective zones, with the corresponding labels for the dominant burning phase. The solid line above the envelope denotes the total mass.}
\label{kip2}
\end{figure}

\noindent Convective cores and shells (higher mass coordinate) for all burning stages except silicon were studied. A 15M$_\odot$, solar metallicity (Z=0.014) stellar model computed using GENEC was evolved up to the end of oxygen burning. Snapshots of the structure were analysed at the start; maximum mass extent; and the end of each burning phase (H, He, C, Ne and O) for both core and shell burning regions.\\

\noindent Fig. \ref{kip2} shows the mass structure including convective instabilities over the evolution of the star. 

\begin{figure}[h!]
\centering
\includegraphics[width=\textwidth]{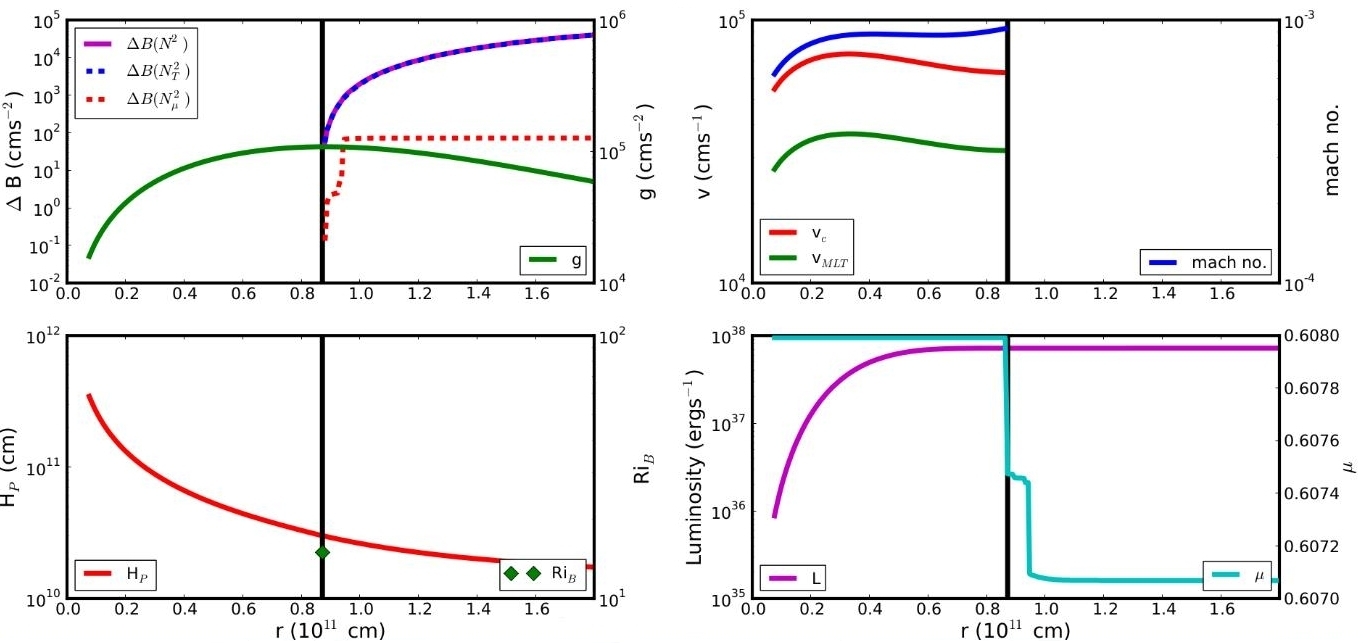}
\caption{Structure properties of the start of core hydrogen burning as a function of radius (0.3\% of fuel burnt).\hspace{\textwidth} \textit{Top left} - Buoyancy jump (magenta) and its components, thermal (blue dashed) and compositional (red dashed) and gravitational acceleration (green).\hspace{\textwidth} \textit{Top right} - Convective (red), mixing length theory (green) velocities and Mach number (blue).\hspace{\textwidth} \textit{Bottom left} - Pressure scale height (red) and bulk Richardson number (green diamond).\hspace{\textwidth} \textit{Bottom right} - Luminosity (magenta) and mean molecular weight (cyan). Vertical black lines represent radial positions of convective boundaries.}
\label{hcores}
\end{figure}

\noindent The thermodynamic structure over the boundaries is continuous. However, there is a discontinuity in composition at the boundary, acting as a `stabilising wall' against convective elements, most easily seen as a positive entropy gradient. These features can be seen in Fig. \ref{hcores}, showing variables at the start of core H-burning. Notable features are: the build up of a molecular weight gradient, due to the migration of the convective boundary; and the jump in buoyancy within the stable region which is dominated by the thermal component.\\ 

\begin{figure}[h!]
\centering
\includegraphics[width=\textwidth]{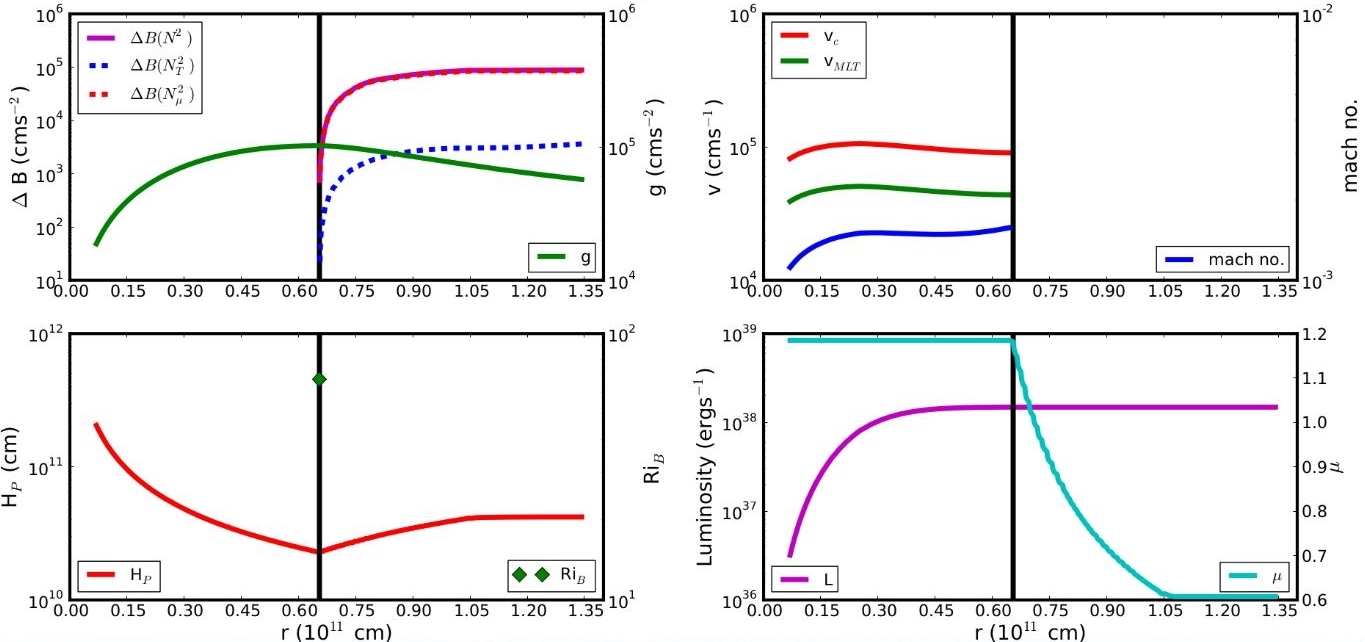}
\caption{Structure properties of the end of core hydrogen burning as a function of radius (90.5\% of fuel burnt). Curves, colours and lines are the same as in Fig. \protect\ref{hcores}.}
\label{hcoree}
\end{figure}

\noindent At the end of core H-burning (Fig. \ref{hcoree}) the molecular weight has risen due to a He rich core, leaving behind a large molecular weight gradient. The recession of the convective core during H-burning leaves behind this signature of mixing, seen through the molecular weight gradient. The jump in buoyancy across the boundary is due mainly to this large compositional gradient, the buoyancy jump acts as a barrier against `convective overshooting'. The magnitude of the jump also increased during the evolution of the core, stiffening the boundary as shown by an increase in the value of the bulk Richardson number.\\

\noindent We are preparing ILESs of the carbon shell in massive stars. We have chosen to simulate the carbon shell for a number of reasons. Firstly, the dominant cooling process is through neutrino losses, this allows thermal diffusion to be neglected, simplifying the computation. Secondly, carbon shell burning is at a mass co-ordinate which coincides with that of the large He-burning core, the compositional history is therefore simpler than the more advanced stages. Lastly, these simulations will be valuable to the community as the carbon shell of massive stars has never been studied in this context before.\\

\begin{figure}[h!]
\centering
\includegraphics[width=\textwidth]{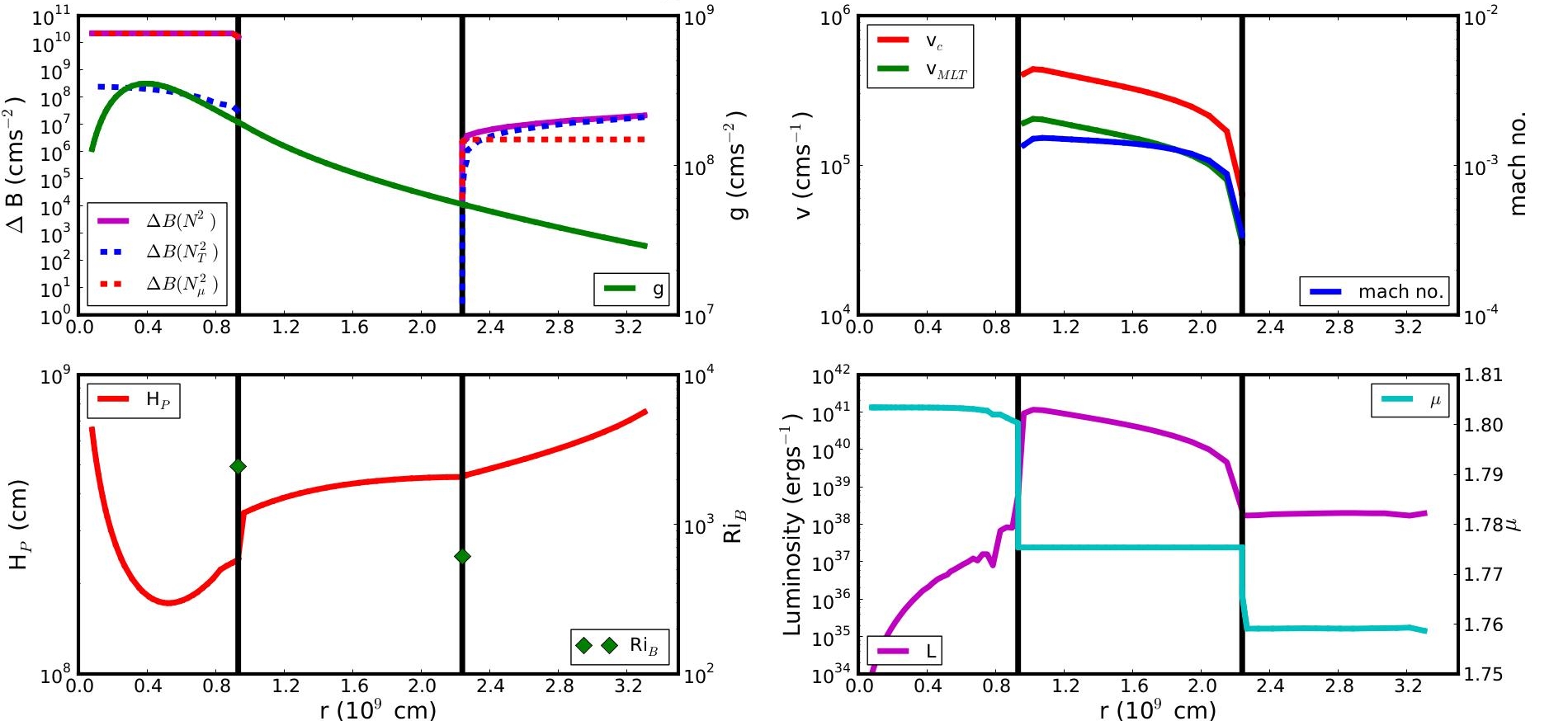}
\caption{Structure properties of carbon shell burning as a function of radius (31\% of the shell's lifetime). Central regions between vertical black lines are convective, areas outside of these lines are stable. Curves, colours and lines are the same as in Fig. \protect\ref{hcores}.}
\label{cshell}
\end{figure}

\noindent In preparation for these simulations, we conducted a parameter study on the second carbon shell (shown in Fig. \ref{kip2}) at mass coordinate $\sim$ 2M$_\odot$. Properties of the shell have been plotted against radius and can be seen in Fig. \ref{cshell}. The defining feature that we found for all convective shells was that the lower boundary is consistently `stiffer' than the corresponding upper boundary, evidenced by the buoyancy jump and the bulk Richardson number. The data is in agreement with 3D simulations of the oxygen burning shell by \citet{2007ApJ...667..448M}. This implies that convective boundary mixing or entrainment is limited at the lower boundary compared to that of the upper boundary. This has important consequences for processes dependent on shallow interior convection such as the onset of novae \citep{2013ApJ...762....8D} and also flame front propagation in S-AGB stars (\citealt{2013ApJ...772..150J,2013ApJ...772...37D}).\\

\begin{figure}[h!]
\centering
\includegraphics[width=\textwidth]{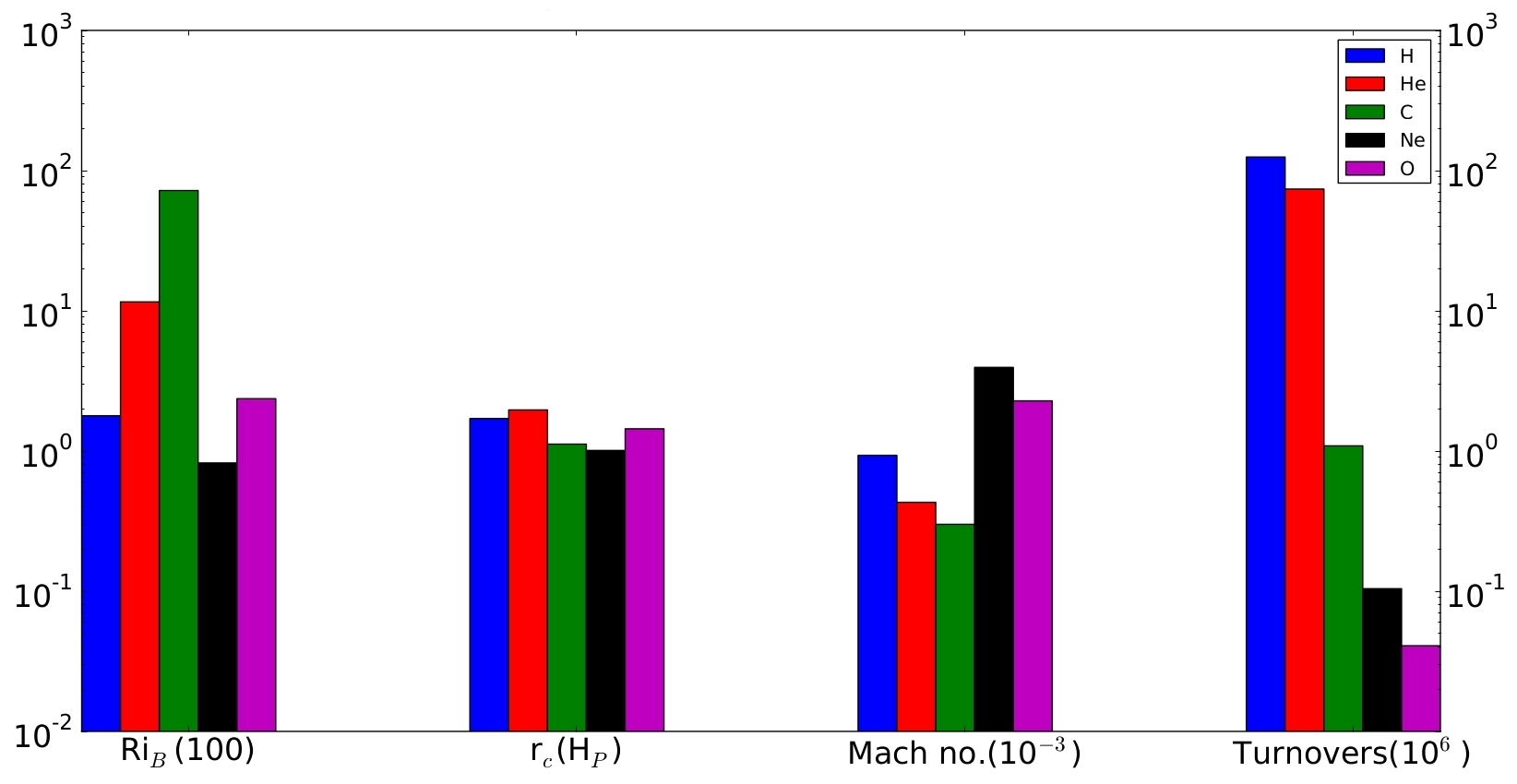}
\caption{Initial core properties at different burning phases: hydrogen (blue); helium (red); carbon (green); neon (black); and oxygen (magenta). From left to right along the horizontal axis are: the bulk Richardson number (normalised by 100); core radius normalised by the mass averaged pressure scale height in the core; mass averaged Mach number within the core (normalised by 10$^{-3}$); and the estimated number of turnovers for the entire phase (normalised by 10\;$^6$).}
\label{hist1}
\end{figure}

\begin{figure}[h!]
\centering
\includegraphics[width=\textwidth]{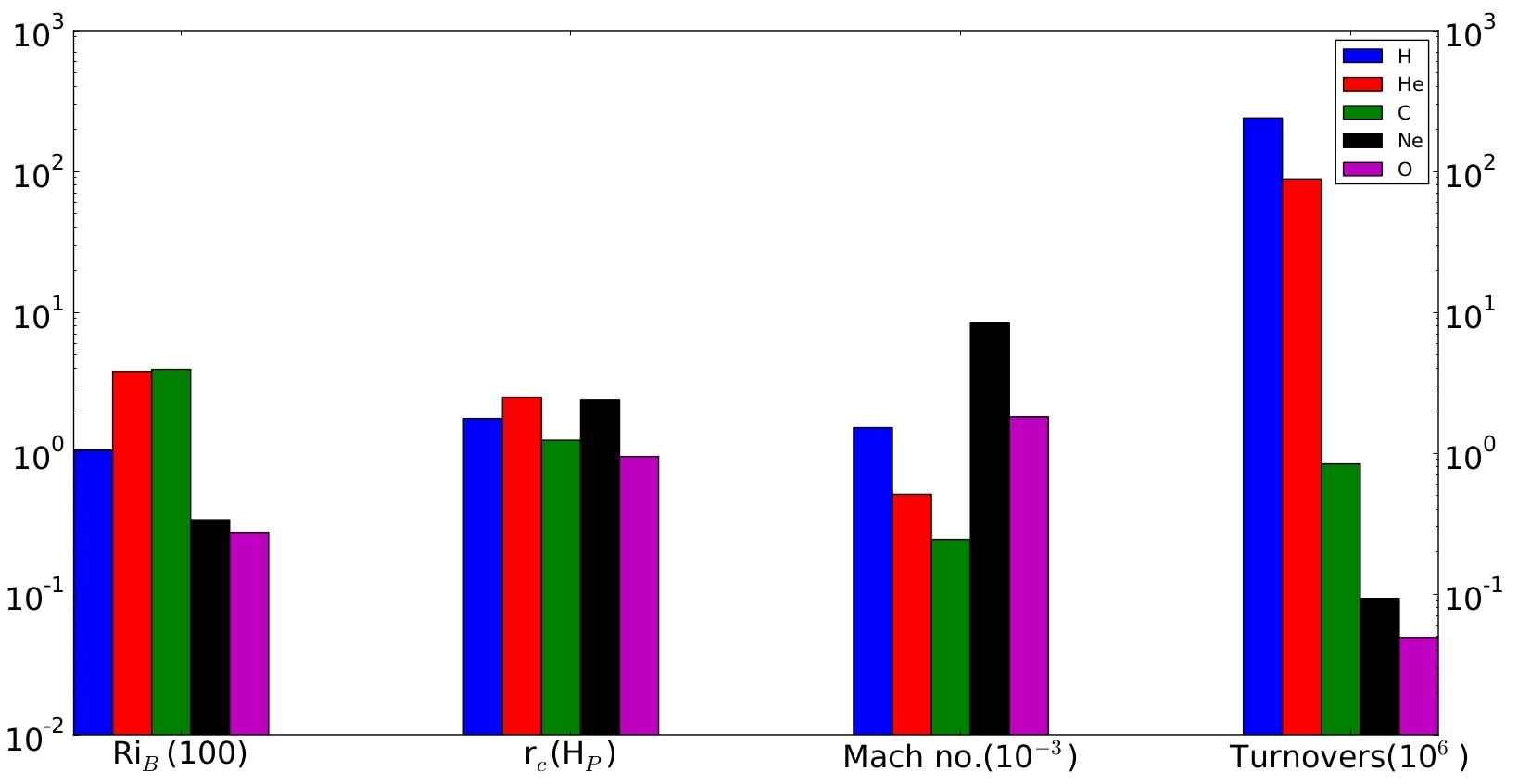}
\caption{Final core properties at different burning phases. Fields, colours and labels are the same as in Fig. \protect\ref{hist1}.}
\label{hist2}
\end{figure}

\noindent Convective regions and boundaries of a 15M$_\odot$ stellar model were analysed. Variables were calculated during different phases of the stellar model, these are detailed in Table \ref{tab3} and Figs. \ref{hist1} and \ref{hist2}. The bulk Richardson number was calculated using Eq. \ref{ribeq}, with $\Delta r=0.1H_{P,0}$ for convective cores and $\Delta r=0.05H_{P,0}$ for convective shells, where $H_{P,0}$ is the pressure scale height on the stable side of the boundary. The peak value of the luminosity $L$ is given in Table \ref{tab3}. The radius of the convective region $r_c$ normalised by the mass averaged pressure scale height over the convective region is also presented. The convective velocity is a mass weighted RMS average:\\

\begin{equation}
v_{c,avg}=\sqrt{\frac{1}{m_2-m_1}\int_{m_1}^{m_2}v_c^2\;dm}.
\end{equation}

\noindent The gravitational acceleration, pressure scale height, mean molecular weight and Mach number have all been mass averaged over the convective region. The number of turnovers for each phase is an estimate based on the following formula:\\

\begin{equation}
\textrm{no. of turnovers}=\frac{\tau_{conv}}{\tau_{burn}}=\frac{2r_{c}/v_{c,avg}}{\tau_{burn}},
\end{equation}

\noindent where $r_{c}$ is the radial extent of the convective zone and $\tau_{burn}$ is the lifetime of each phase. The numerical values for the bulk Richardson number (Ri$_B$) presented in Table \ref{tab3} are to be taken with caution and only as estimates because these were determined from a 1D stellar model. Estimates were used for the turbulent length scale, turbulent velocity and integral scale for the calculation of Ri$_B$. It is nevertheless interesting to compare values of Ri$_B$ between different phases of a burning stage, for example, for the carbon and oxygen-burning convective cores. In these two cases, Ri$_B$ is the highest during the maximum extent of the convective zone and lowest towards the end of the burning stage. This can be understood by the fact that between the start and maximum extent of these burning stages, a mean molecular weight gradient gradually builds up at the upper boundary. At the end of a burning stage, convective zones tend to recede and thus the compositional gradient at the edge of the convective zone is reduced due to prior convective mixing. The values for Ri$_B$ confirm that lower boundaries of convective shells are stiffer than the upper boundaries.

\begin{sidewaystable}
\begin{center}
\begin{tabulary}{\textwidth}{l c c c c c c c}
\hline \hline\\
\textbf{Phase} & \textbf{Ri$_B$} & \textbf{g$_{avg}$ (cm$\,s^{-2}$)} & \textbf{v$_{c,avg}$ (cm$\,s^{-1}$)} & \textbf{L$_{peak}$ (erg$\,s^{-1}$)} & \textbf{H$_{P,avg}$ (cm)} & $\boldsymbol{\mu}_{avg}$ & \textbf{Ma$_{avg}$}\\\\
\hline\hline\\
\textbf{ H Core Start } &  178  &  8.7$\times10^{4}$  &  6.9$\times10^{4}$  &  7.5$\times10^{37}$  &  5.1$\times10^{10}$  &  0.627  &  9.3$\times10^{-4}$ \\
\textbf{ H Core End } &  105  &  9.0$\times10^{4}$  &  9.7$\times10^{4}$  &  1.5$\times10^{38}$  &  3.6$\times10^{10}$  &  1.21  &  1.5$\times10^{-3}$ \\\\
\hline\\
\textbf{ He Core Start } &  1160  &  1.8$\times10^{6}$  &  4.7$\times10^{4}$  &  4.6$\times10^{37}$  &  5.1$\times10^{9}$  &  1.36  &  4.3$\times10^{-4}$ \\
\textbf{ He Core End } &  381  &  2.4$\times10^{6}$  &  5.9$\times10^{4}$  &  1.2$\times10^{38}$  &  4.2$\times10^{9}$  &  1.71  &  5.1$\times10^{-4}$ \\\\
\hline\\
\textbf{ C Core Start } &  7210  &  4.3$\times10^{7}$  &  6.9$\times10^{4}$  &  5.9$\times10^{38}$  &  8.9$\times10^{8}$  &  1.77  &  3.0$\times10^{-4}$ \\
\textbf{ C Core Max } &  1.2$\times10^{4}$  &  4.6$\times10^{7}$  &  5.6$\times10^{4}$  &  5.6$\times10^{38}$  &  8.3$\times10^{8}$  &  1.78  &  2.4$\times10^{-4}$ \\
\textbf{ C Core End } &  393  &  5.0$\times10^{7}$  &  5.8$\times10^{4}$  &  6.1$\times10^{38}$  &  8.8$\times10^{8}$  &  1.87  &  2.4$\times10^{-4}$ \\\\
\hline\\
\textbf{ Ne Core Start } &  82  &  4.8$\times10^{8}$  &  1.5$\times10^{6}$  &  1.7$\times10^{43}$  &  2.3$\times10^{8}$  &  1.81  &  3.9$\times10^{-3}$ \\
\textbf{ Ne Core Max } &  52  &  4.1$\times10^{8}$  &  2.6$\times10^{6}$  &  1.4$\times10^{44}$  &  2.1$\times10^{8}$  &  1.83  &  7.3$\times10^{-3}$ \\
\textbf{ Ne Core End } &  33  &  4.2$\times10^{8}$  &  3.0$\times10^{6}$  &  2.1$\times10^{44}$  &  2.1$\times10^{8}$  &  1.83  &  8.3$\times10^{-3}$ \\\\
\hline\\
\textbf{ O Core Start } &  236  &  4.5$\times10^{8}$  &  8.8$\times10^{5}$  &  5.2$\times10^{42}$  &  2.3$\times10^{8}$  &  1.84  &  2.2$\times10^{-3}$ \\
\textbf{ O Core Max } &  8.5$\times10^{4}$  &  5.0$\times10^{8}$  &  7.9$\times10^{5}$  &  7.4$\times10^{42}$  &  2.0$\times10^{8}$  &  1.86  &  2.0$\times10^{-3}$ \\
\textbf{ O Core End } &  27  &  5.1$\times10^{8}$  &  7.5$\times10^{5}$  &  3.1$\times10^{42}$  &  2.5$\times10^{8}$  &  1.91  &  1.8$\times10^{-3}$ \\\\
\hline\\
\textbf{ He Shell Start } &  1130  ( 606 ) &  7.4$\times10^{7}$  &  5.2$\times10^{4}$  &  5.2$\times10^{39}$  &  5.5$\times10^{8}$  &  1.76  &  2.5$\times10^{-4}$ \\
\textbf{ He Shell End } &  307  ( 1.9$\times10^{4}$ ) &  7.2$\times10^{7}$  &  4.7$\times10^{4}$  &  1.6$\times10^{39}$  &  6.3$\times10^{8}$  &  1.74  &  2.8$\times10^{-4}$ \\\\
\hline\\
\textbf{ C Shell Start } &  1440  ( 2.0$\times10^{4}$ ) &  2.3$\times10^{8}$  &  1.9$\times10^{5}$  &  2.0$\times10^{41}$  &  2.5$\times10^{8}$  &  1.8  &  7.5$\times10^{-4}$ \\
\textbf{ C Shell End } &  1600  ( 1.2$\times10^{5}$ ) &  2.1$\times10^{8}$  &  1.2$\times10^{5}$  &  3.3$\times10^{40}$  &  2.9$\times10^{8}$  &  1.8  &  4.8$\times10^{-4}$ \\\\
\hline\\
\textbf{ O Shell Start } &  1.0$\times10^{5}$  ( 1.1$\times10^{5}$ ) &  1.3$\times10^{9}$  &  9.1$\times10^{4}$  &  1.6$\times10^{41}$  &  1.2$\times10^{8}$  &  1.88  &  2.0$\times10^{-4}$ \\
\textbf{ O Shell End } &  1100  ( 1.6$\times10^{5}$ ) &  1.5$\times10^{9}$  &  2.5$\times10^{5}$  &  3.7$\times10^{42}$  &  1.0$\times10^{8}$  &  1.92  &  6.2$\times10^{-4}$ \\\\
\hline\hline
\end{tabulary}
\caption{Bulk Richardson number, gravitational acceleration ($cm\,s^{-2}$), convective velocity ($cm\,s^{-1}$), luminosity ($erg\,s^{-1}$), pressure scale height (cm), mean molecular weight and Mach number of different times during core and shell burning phases of a 15M$_\odot$ stellar model. Bulk Richardson numbers are boundary values, brackets indicate values at the lower boundary. Except for the luminosity which is the peak value, all other values were mass averaged over the convective region.}
\label{tab3}
\end{center}
\end{sidewaystable}

\clearpage

\clearpage

\section{Conclusion}

\noindent 1D stellar evolution models are well equipped to simulate the entire evolution of massive stars, but lack detail in the treatment of convective energy transport and turbulent mixing. 3D hydrodynamic simulations, specifically ILES can give detailed information about stellar interiors, which can provide useful insight into new prescriptions for stellar models.\\

\noindent In this paper, after reviewing the main aspects of 1D and 3D modelling, we presented a parameter study on a 15 M$_\odot$ model, for all convective regions including both the radiation dominated early phases and the neutrino-cooled advanced phases. We present their fluid properties in terms of quantities relevant to prepare 3D simulations of these phases.\\ 

\noindent The summary of our main findings are as follows. During the initial growth stage of the convective region the buoyancy jump over the boundary is thermally dominated. As the burning stage progresses the convective boundary recedes, a compositional gradient is built up and the buoyancy jump becomes gradually dominated by a molecular weight gradient. The ratio of the work against buoyancy to the turbulent kinetic energy is a measure of the boundary `stiffness', i.e. the bulk Richardson number. The value of this ratio is relatively large at the lower radial boundary of convective shells. The ratio at the upper boundary is smaller, allowing more material to be entrained over the region's evolution.\\ 

\noindent Future plans include a resolution study of an idealised hydrodynamic simulation of the carbon shell, following this an additional parameter study in 3D and finally a high-resolution, realistic setup run over multiple turnover times. Using a Reynolds-Averaged Navier-Stokes framework to spatially average data along a radial axis (\citealt{2013ApJ...769....1V}), diffusion coefficients will be extracted and compared to those obtained from 1D stellar models. Another important phase to study is the possible late shell mergers occurring after core silicon burning.\\

\noindent By simulating different epochs of the carbon shell burning phase, improved or replacement parameterisations for additional convective mixing beyond the conventional Schwarzschild or Ledoux criteria can be formulated. With more accurate mixing prescriptions, stellar models will better describe the structure and evolution of massive stars in general, as well as changing the final nucleosynthesis yields. Also, from the point of view of other communities dealing with turbulent mixing processes, turbulence in such extreme conditions ($\textrm{Re} > 10^8$) will be better understood, in addition to similar studies on stellar interiors (e.g. \citealt{2007ApJ...667..448M}; and further work since this workshop by \citealt{camp2015}). \\

\noindent The authors acknowledge support from EU-FP7-ERC-2012-St Grant 306901. RH acknowledges support from the World Premier International Research Center Initiative (WPI Initiative), MEXT, Japan. This work used the Extreme Science and Engineering Discovery Environment (XSEDE), which is supported by National Science Foundation grant number OCI-1053575. CM and WDA acknowledge support from NSF grant 1107445 at the University of Arizona. MV acknowledges support from the European Research Council through grant ERC-AdG No. 341157-COCO2CASA. This work used the DiRAC Data Centric system at Durham University, operated by the Institute for Computational Cosmology on behalf of the STFC DiRAC HPC Facility (www.dirac.ac.uk). This equipment was funded by BIS National E-infrastructure capital grant ST/K00042X/1, STFC capital grants ST/H008519/1 and ST/K00087X/1, STFC DiRAC Operations grant ST/K003267/1 and Durham University. DiRAC is part of the National E-Infrastructure.

\clearpage
\bibliography{references}

\end{document}